\documentclass[aps, 10pt, prd,notitlepage,twocolumn, superscriptaddress]{revtex4-2}

\usepackage{amsmath,amssymb,amsfonts}
\usepackage[utf8]{inputenc}
\usepackage[T1]{fontenc}
\usepackage{mathrsfs}
\usepackage[colorlinks=true]{hyperref}
\hypersetup{citecolor=cyan,linkcolor=magenta}
\usepackage{anyfontsize}
\usepackage[usenames,dvipsnames]{xcolor}

\usepackage{newtxtext}
\usepackage{newtxmath}
\usepackage{tensor}
\usepackage{bm}

\usepackage{graphicx}

\usepackage{natbib}

\usepackage[normalem]{ulem}

\def\jnl@style{\it}
\def\aaref@jnl#1{{\jnl@style#1}}

\def\aaref@jnl#1{{\jnl@style#1}}

\def\aj{\aaref@jnl{AJ}}                   
\def\apj{\aaref@jnl{ApJ}}                 
\def\apjl{\aaref@jnl{ApJ}}                
\def\apjs{\aaref@jnl{ApJS}}               
\def\apss{\aaref@jnl{Ap\&SS}}             
\def\aap{\aaref@jnl{A\&A}}                
\def\aapr{\aaref@jnl{A\&A~Rev.}}          
\def\aaps{\aaref@jnl{A\&AS}}              
\def\mnras{\aaref@jnl{Mon.~Not.~Roy.~Astron.~Soc.}}             
\def\prd{\aaref@jnl{Phys.~Rev.~D}}        
\def\prc{\aaref@jnl{Phys.~Rev.~C}}  
\def\prl{\aaref@jnl{Phys.~Rev.~Lett.}}    
\def\qjras{\aaref@jnl{QJRAS}}             
\def\skytel{\aaref@jnl{S\&T}}             
\def\ssr{\aaref@jnl{Space~Sci.~Rev.}}     
\def\zap{\aaref@jnl{ZAp}}                 
\def\nat{\aaref@jnl{Nature}}              
\def\aplett{\aaref@jnl{Astrophys.~Lett.}} 
\def\apspr{\aaref@jnl{Astrophys.~Space~Phys.~Res.}} 
\def\physrep{\aaref@jnl{Phys.~Rep.}}      
\def\physscr{\aaref@jnl{Phys.~Scr}}       
\def\commat{\aaref@jnl{Comm.~Math.~Phys.}}              
\def\science{\aaref@jnl{Science}}               
\def\cqg{\aaref@jnl{Classical Quant.~Grav.}}            
\def\jpcs{\aaref@jnl{JPCS}}                                     
\def\ijmpd{\aaref@jnl{Int.~J.~Mod.~Phys.~D}}                    
\def\grg{\aaref@jnl{Gen.~Relat.~Gravit.}}               
\def\rpp{\aaref@jnl{Rep.~Prog.~Phys.}}          
\def\npa{\aaref@jnl{Nucl.~Phys.~A}}        
\def\lrr{\aaref@jnl{Living Rev.~Rel.}}                   
\def\jcap{\aaref@jnl{J.~Cosmology Astropart.~Phys.}}    
\def\rmp{\aaref@jnl{Rev.~Mod.~Phys.}}   

\allowdisplaybreaks[1]

\begin{document}

\title{Dynamical  descalarization with a jump during black hole merger}

\author{Daniela D. Doneva}
\affiliation{Theoretical Astrophysics, Eberhard Karls University of T\"ubingen, T\"ubingen 72076, Germany}

\author{Alex Va\~n\'o-Vi\~nuales}
\affiliation{CENTRA, Departamento de F\'isica, Instituto Superior T\'ecnico IST, Universidade de Lisboa UL, Avenida Rovisco Pais 1, 1049-001 Lisboa, Portugal}

\author{Stoytcho S. Yazadjiev}
\affiliation{Theoretical Astrophysics, Eberhard Karls University of T\"ubingen, T\"ubingen 72076, Germany}
\affiliation{Department of Theoretical Physics, Faculty of Physics, Sofia University, Sofia 1164, Bulgaria}
\affiliation{Institute of Mathematics and Informatics, 	Bulgarian Academy of Sciences, 	Acad. G. Bonchev St. 8, Sofia 1113, Bulgaria}

\date{\today}

\begin{abstract}
\noindent Black holes in scalar-Gauss-Bonnet gravity are prone to scalarization, that is a spontaneous development of scalar hair for strong enough spacetime curvature while the weak field regime of the theory coincides with general relativity. Since large spacetime curvature is associated with smaller black hole masses, the merging of black holes can lead to dynamical descalarization. This is a spontaneous release of the scalar hair of the newly formed black hole in case its mass is above the scalarization threshold. Depending on the exact form of the Gauss-Bonnet coupling function, the stable scalarized solutions can be either continuously connected to the Schwarzschild black hole, or the transitions between the two can happen with a jump. By performing simulations of black hole head-on collisions in  scalar-Gauss-Bonnet gravity prone to dynamical descalization, we have demonstrated that such a jump can be clearly observed in the accumulated gravitational wave data of multiple merger events with different masses. The distinct signature in the gravitational wave signal will share similarities with the effects expected from first order matter phase transitions happening during neutron star binary mergers.
\end{abstract}

\maketitle

\section{Introduction.}
It has been an important goal for decades to probe the strong-field regime of gravity and to see whether the predictions of general relativity (GR) work on that scale as well. The direct detection of gravitational waves and the subsequent rapid advance of the gravitational wave detectors give us hope that soon gravity will be better understood in the realm of large spacetime curvature. The progress in observations calls for further advances in modeling phenomena in modified theories of gravity that lead to strong gravitational wave emission. Among the best candidates in this respect are binary black hole mergers \cite{Baiotti:2016qnr}. Due to their complexity, the numerical merger simulations in modified gravity were performed only in a handful of cases \cite{Healy:2011ef,Okounkova:2019dfo,Okounkova:2019zjf,Okounkova:2020rqw,East:2021bqk,Silva:2020omi}. They demonstrated, though, the big potential in using merger events for constraining alternative theories of gravity.

There is a whole plethora of attempts to generalize Einstein's theory of gravity but many, even though very promising on certain scales, struggle to reproduce a weak field limit that is in agreement with the observation \cite{Will14,Berti:2015itd}. In order to reconcile gravitational theory predictions on different scales, often different types of screening mechanisms are introduced. A class of theories, offering a natural build-in screening mechanism are the theories of gravity allowing for the development of the so-called spontaneous scalarization \cite{Damour92}. These are classes of extended scalar-tensor theories admitting always the GR zero scalar field solutions, but allowing for the development of a scalar field in the realm of strong spacetime curvature. The spontaneous scalarization can be considered as a phase transition from a GR-like state to a scalarized one that is normally energetically favorable. 

Initially developed in the framework of neutron stars \cite{Damour92}, it was recently shown that black holes can also scalarize in scalar-Gauss-Bonnet (sGB) gravity \cite{Doneva:2017bvd,Silva:2017uqg,Antoniou:2017acq} that made them an interesting candidate for exploring astrophysical implications (see \cite{Stefanov:2007eq,Doneva:2010ke} for the first results on scalarized charged black holes). The stability and quasinormal modes of such black holes was examined in \cite{Blazquez-Salcedo:2018jnn,Blazquez-Salcedo:2020rhf,Blazquez-Salcedo:2020caw,Staykov:2021dcj}. Subsequently, the topic was largely developed to include different types of coupling \cite{Doneva:2018rou,Silva:2018qhn,Minamitsuji:2018xde}, scalar field potential \cite{Macedo:2019sem,Doneva:2019vuh}, rapid rotation \cite{Cunha:2019dwb,Collodel:2019kkx}, spin-induced scalarization \cite{Dima:2020yac,Doneva:2020nbb,Berti:2020kgk,Herdeiro:2020wei,Doneva:2020kfv,Annulli:2022ivr}. Scalarized neutron stars in sGB gravity were considered in \cite{Silva:2017uqg,Doneva:2017duq,Staykov:2021jfi,Ventagli:2021ubn,Xu:2021kfh} and they were confronted against the binary pulsar experiments leading to strong constraints on the theory parameters \cite{Danchev:2021tew}. It was also demonstrated that black holes can scalarize in a wider variety of extended scalar-tensor theories \cite{Andreou:2019ikc,Ventagli:2020rnx,Antoniou:2021zoy,Antoniou:2022agj} and the case of Einstein-Maxwell-scalar gravity was examined in \cite{Herdeiro:2018wub,Fernandes:2019rez,Herdeiro:2021vjo}.

Even though interesting both from theoretical and astrophysical points of view, the introduction of the Gauss-Bonnet (GB) invariant brings a great complication compared to GR especially when considering the compact object dynamics. That is why the advance in this direction is still limited. The dynamics of the scalar field around isolated black holes was considered in \cite{Benkel:2016kcq,Ripley:2019aqj,Ripley:2020vpk,East:2020hgw,Doneva:2021dqn,Kuan:2021lol,Doneva:2021tvn}. Binary black hole merger in GB gravity was studied in the decoupling limit \cite{Silva:2020omi}, i.e. when the scalar field backreaction on the metric is neglected. Dynamical descalarization, which is the spontaneous ``release'' of the scalar field, was observed after the merger when one or both of the individual black holes fulfilled the criteria for scalar field development. However, the mass of the newly formed black hole was beyond the point of scalarization. The full problem without approximation was simulated in \cite{East:2021bqk} where more light was shed on the loss of hyperbolicity for such systems. The formation of scalarized compact objects through stellar core collapse in sGB theory was considered in \cite{Kuan:2021lol}. 

All studies of black hole scalarization dynamics in sGB gravity until now considered the standard case when the stable scalarized black hole branch is continuously connected to the GR one. Allowing for a more general form of the coupling can easily cause a part of the scalarized branch to lose stability and effectively create a jump between the last stable scalarized solution and the GR branch, despite a continuous connection existing between both. Even more, a region in the parameter space can exist where both the Schwarzschild solution and the scalarized ones are stable. The properties of such static black holes were considered in \cite{Doneva:2021tvn} and their stability was examined later in \cite{Blazquez-Salcedo:2022omw}. Similar behavior is observed also in other modified theories of gravity for charged black holes \cite{Blazquez-Salcedo:2020nhs,LuisBlazquez-Salcedo:2020rqp}. In the present paper we study their astrophysical implications by simulating the merger of two black holes in sGB gravity for a coupling function allowing for such a jump and discuss the observational consequences. It is interesting to note that the presence of a gap between two stable branches of solutions is also observed for neutron stars possessing first-order matter phase transition from confined hadronic to deconfined quark matter \cite{kam81,Glendenning:1998ag,sch00,shaf02} and even though we consider here quite distinct compact objects, the merger will possess certain clear similarities with \cite{Most:2018eaw,Bauswein:2018bma,Weih:2019xvw}.

The paper is organized as follows. In Sec. II we present the basic formalism and the properties of the isolated black holes under consideration. In the next Sec. III, we discuss the 3+1 decomposition of the scalar field equation and the numerical approach. The binary black hole merger results are presented in Sec. IV. The paper ends with a discussion.

\section{Scalarized black holes  in Gauss-Bonnet gravity}
The action in scalar-Gauss-Bonnet gravity has the following form:  	
\begin{eqnarray}\label{GBA}
S=&&\frac{1}{16\pi}\int d^4x \sqrt{-g} 
\Big[R - 2\nabla_\mu \varphi \nabla^\mu \varphi 
+ \lambda^2 f(\varphi){\cal R}^2_{GB} \Big] .\label{eq:quadratic}
\end{eqnarray}
where $R$ is the Ricci scalar with respect to the spacetime metric $g_{\mu\nu}$, $\varphi$ denotes the scalar field, $f(\varphi)$ is the coupling between the scalar field and the Gauss-Bonnet invariant, $\lambda$ is the so-called Gauss-Bonnet coupling constant having  dimension of $length$ and ${\cal R}^2_{GB}=R^2 - 4 R_{\mu\nu} R^{\mu\nu} + R_{\mu\nu\alpha\beta}R^{\mu\nu\alpha\beta}$. 

Varying the action with respect to the metric and the scalar field results in the following system of field equations:
\begin{eqnarray}\label{FE}
&&R_{\mu\nu}- \frac{1}{2}R g_{\mu\nu} + \Gamma_{\mu\nu}= 2\nabla_\mu\varphi\nabla_\nu\varphi -  g_{\mu\nu} \nabla_\alpha\varphi \nabla^\alpha\varphi - \frac{1}{2} g_{\mu\nu}V(\varphi),\\
&&\nabla_\alpha\nabla^\alpha\varphi= \frac{1}{4} \frac{dV(\varphi)}{d\varphi} -  \frac{\lambda^2}{4} \frac{df(\varphi)}{d\varphi} {\cal R}^2_{GB},
\end{eqnarray}
where $\Gamma_{\mu\nu}$ is defined by 
\begin{eqnarray}
\Gamma_{\mu\nu}&=& - R(\nabla_\mu\Psi_{\nu} + \nabla_\nu\Psi_{\mu} ) - 4\nabla^\alpha\Psi_{\alpha}\left(R_{\mu\nu} - \frac{1}{2}R g_{\mu\nu}\right) \\
&& + 4R_{\mu\alpha}\nabla^\alpha\Psi_{\nu} + 4R_{\nu\alpha}\nabla^\alpha\Psi_{\mu} \nonumber \\ 
&& - 4 g_{\mu\nu} R^{\alpha\beta}\nabla_\alpha\Psi_{\beta} 
 + \,  4 R^{\beta}_{\;\mu\alpha\nu}\nabla^\alpha\Psi_{\beta} .
\end{eqnarray}  
Here $\Psi_{\mu}$ encodes the coupling between the scalar field and the GB invariant and we have defined it as:
\begin{eqnarray}
\Psi_{\mu}= \lambda^2 \frac{df(\varphi)}{d\varphi}\nabla_\mu\varphi .
\end{eqnarray}

In what follows for simplicity we will assume that the scalar field potential is zero. The remaining freedom we have within the sGB gravity under consideration is the choice of $f(\varphi)$.
In order to have spontaneous (de)scalarization we must require that $(df/d\varphi)(\varphi=0)=0$ (which ensures that the GR solutions are also  solutions of the field equations \eqref{FE})   and $(df^2/d\varphi^2)(\varphi=0)\ne 0$.  By taking the scalar field equation in \eqref{FE} and perturbing the scalar field around the background scalar field value $\varphi_0=0$ \footnote{One can easily assume a nonzero $\varphi_0$ that remains the formalism below unchanged.} on a fixed GR background,  one can reach the following expression:
\begin{equation}\label{FE_perturb}
 \nabla_\alpha\nabla^\alpha \delta \varphi= -  \frac{\lambda^2}{4} \frac{d^2f(\varphi)}{d\varphi^2} {\cal R}^2_{GB}.
\end{equation}
Through this equation an effective scalar field mass can be defined 
\begin{equation}
	\mu_{\rm eff}^2 = -\frac{\lambda^2}{4}\left.\frac{df^2}{d\varphi^2}\right|_{\varphi=0} {\cal R}_{GB}^2.
\end{equation}
For $\mu_{\rm eff}^2<0$ a tachyon instability will be present  driving the scalar field away from the trivial value and endowing the compact object with a scalar field. For Schwarzschild black hole ${\cal R}_{GB}^2>0$ always and a negative effective mass requires $(df^2/d\varphi^2)(\varphi=0)>0$.

The simplest function satisfying these conditions is  $f(\varphi)=\varphi^2$. It leads, though, to unstable scalarized black hole solutions \cite{Blazquez-Salcedo:2018jnn}. One can go one step further and add a quartic scalar field term to the coupling that can potentially stabilize the solutions \cite{Silva:2018qhn,Minamitsuji:2018xde}. From a numerical perspective, it has proven much more convenient to work with a coupling being an exponential function of the scalar field. A common choice is $f(\varphi)=\frac{1}{2\beta}(1-\exp(-\beta \varphi^2))$ \cite{Doneva:2017bvd} that leads to stable and well behaved solutions from a numerical point of view (see e.g. \cite{Cunha:2019dwb,Herdeiro:2020wei,Blazquez-Salcedo:2022omw,Doneva:2018rou}). Binary black hole mergers for such types of coupling have already been considered in \cite{Silva:2020omi,East:2021bqk} and the stellar core-collapse in \cite{Kuan:2021lol}.

\begin{figure}
	\includegraphics[scale=0.35]{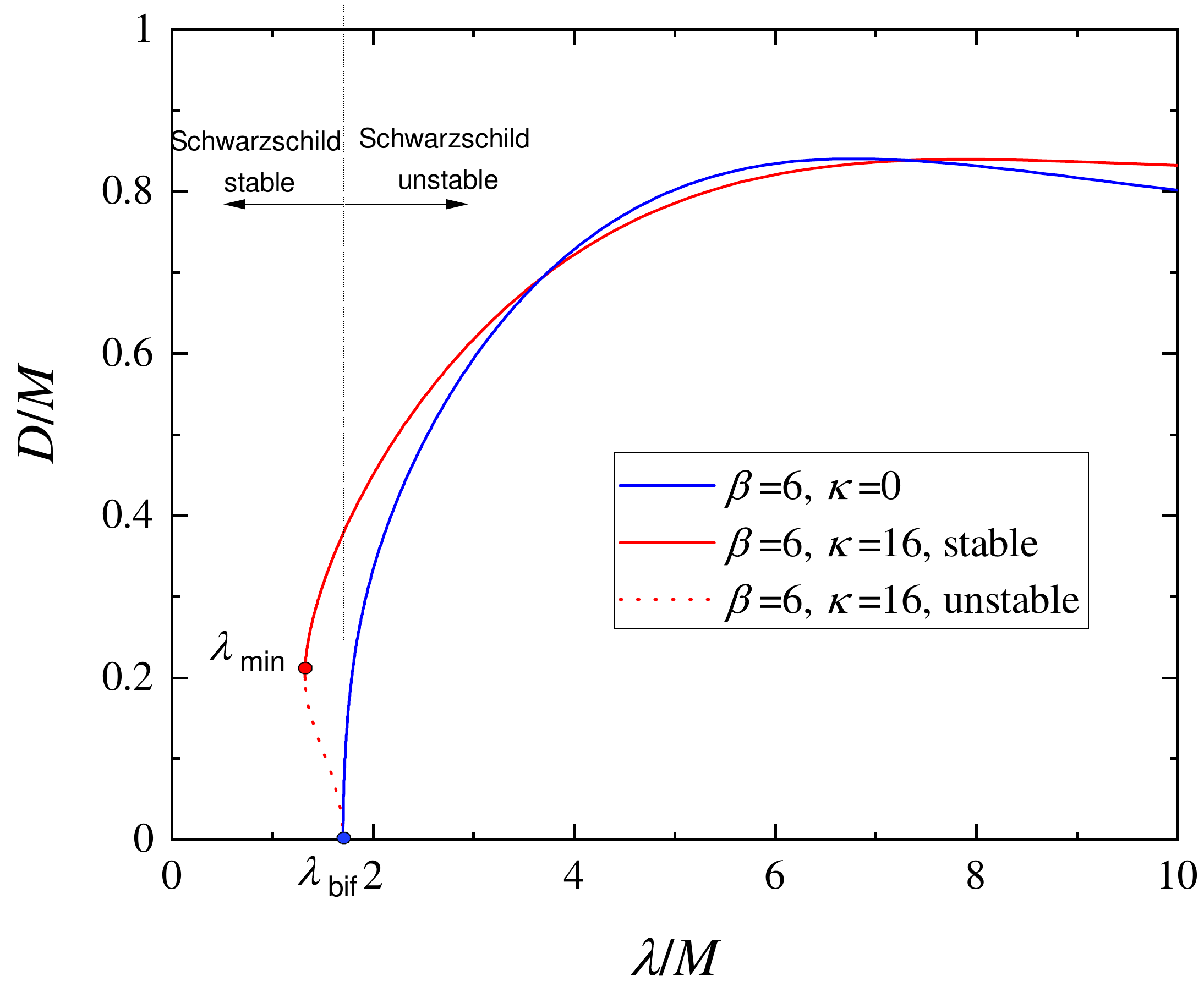}
	\caption{The normalized scalar charge $D/M$ as a function of the normalized GB coupling parameter $\lambda/M$ for sequences of scalarized black holes with $\beta=6$  and two values of $\kappa$. The unstable part of the $\kappa=16$ branch is marked with a dotted line.}
	\label{fig:D_lambda}
\end{figure}

In the present paper we are interested is a slightly different form of the coupling where a quartic scalar field term is added, namely 
\begin{equation}\label{coupling}
	f(\varphi)=\frac{1}{2\beta}(1-\exp(-\beta (\varphi^2 + \kappa \varphi^4))) .
\end{equation}
It has a similar behavior to the above mentioned case with $f(\varphi)=\varphi^2 + \eta \varphi^4$ \cite{Silva:2018qhn,Minamitsuji:2018xde} but we find the exponential function more convenient numerically. For example, the regularity conditions at the horizon are more easily satisfied leading to the existence of scalarized black holes all the way from the bifurcation point to zero black hole mass, see Fig. \ref{fig:D_lambda}. Our tests showed, though, that the effects reported here are generic, qualitatively very similar for a much larger class of couplings, as long as a jump between the GR and sGB black hole branches is present.

The scalar charge as a function of the normalized parameter $\lambda/M$ is shown in Fig. \ref{fig:D_lambda} for  branches of scalarized static black holes with coupling \eqref{coupling}, $\beta=6$ and two values of $\kappa$ . Here the scalar charge $D$ is defined as the leading order asymptotic of the scalar field at infinity, namely $\varphi (r \rightarrow \infty) \sim D/r$. Only the region close to the bifurcation point is shown since this will be the relevant one for the dynamical simulations below. The Schwarzschild black hole in our case is always a solution of the field equations corresponding to the $x$-axis with $D=0$. It destabilizes for $\lambda/M$ larger than the point of bifurcation.  The first choice $\kappa=0$ in the figure is the standard scalarization considered for the first time in \cite{Doneva:2017bvd} where we have the scalarized black holes branching out at a certain $\lambda_{\rm bif}$  with the scalar charge increasing as  $\lambda/M$ increases. This is not the case, though, if $\kappa$ is sufficiently large and the quartic term in the  function \eqref{coupling} starts being important. For the case of $\kappa=16$ in the figure, one can see that after the bifurcation point the branch first moves to the left and, after reaching a minimum $\lambda_{\rm min}$, it turns right. This small portion of the branch after the bifurcation is unstable \cite{Blazquez-Salcedo:2022omw} and thus the last stable scalarized solution is not continuously connected to Schwarzschild. Even more, there is a range of  $\lambda/M$, namely $(\lambda_{\rm min},\lambda_{\rm bif})$, where both the scalarized solutions and the Schwarzschild black hole are linearly stable resembling closely the nonlinear black hole scalarization  \cite{Doneva:2021tvn}. Thus the transition between the two classes of solutions will happen with a jump that will have a very interesting effect on the binary black hole mergers discussed below. 

In the figure we have denoted two points that will be relevant for the further analysis. One is $\lambda_{\rm bif}=1.70M$ that is the point of bifurcation of the scalarized solutions above which Schwarzschild gets unstable. The other one is $\lambda_{\rm min}=1.32M$ that is the minimum value of $\lambda/M$ for which scalarized solutions with $\kappa=16$ exist. As we discussed, in the range $(\lambda_{\rm min},\lambda_{\rm bif})$ the Schwarzschild solutions is stable while black holes with nontrivial scalar field still exist for $\kappa=16$. In the results presented here all quantities are expressed in units of $M$. One can normalize to $\lambda$ instead and then the point of bifurcation will be located at $M_{\rm bif}/\lambda=0.587$ and $\lambda_{\rm min}$ will correspond to a maximum mass for which $\kappa=16$ solutions exist, that is  $M_{\rm max}/\lambda=0.756$.

\section{$3+1$ decomposition and numerical approach}
We will work with the $3+1$ decomposition of the field equations adopting Brown’s covariant form \cite{Brown:2009dd} of the BSSN formalism \cite{NOK,PhysRevD.52.5428,Baumgarte:1998te} in curvilinear coordinates \cite{Bonazzola:2003dm,Shibata:2004qz,Montero:2012yr, Baumgarte:2012xy}. The general form of the spacetime metric in the $3+1$ formalism is
\begin{equation}
    ds^2 = g_{\mu\nu} dx^\mu dx^\nu = -\alpha^2 dt^2 + \gamma_{ij} (dx^i + \beta^i dt)(dx^j + \beta^j dt)
\end{equation}
where $\alpha$ is the lapse function, $\beta^i$ is the shift vector, and $\gamma_{ij}$ is the three-dimensional spacial metric. We introduce a conformally related spatial metric ${\bar \gamma_{ij}}$ via the relation 
\begin{equation}
	{\bar \gamma_{ij}} = e^{-4\phi}\gamma_{ij}
\end{equation}
where the conformal factor is $e^{-4\phi}$. In contrast to the original BSSN formulation, the determinant ${\bar \gamma}$ is not required to be unity, but instead, its time derivative vanishes.

In our simulations, we will adopt the decoupling limit approximation where the backreaction of the scalar field on the spacetime metric is neglected similar to \cite{Silva:2020omi}. Even though this is an approximation, it captures very well the scalar field dynamics and can be even quantitative quite accurate for relatively weak scalar fields \cite{Doneva:2021dqn}. Thus it can perfectly serve our purpose to study the qualitative effect of the presence of a jump in the solution space visible in Fig. \ref{fig:D_lambda}. Within this approximation clearly, Einstein's field equations remain unchanged. The pure GR BSSN formalism in spherical coordinates has been developed in a series of papers \cite{Bonazzola:2003dm,Shibata:2004qz,Brown:2009dd,Montero:2012yr, Baumgarte:2012xy}. Since we neglect the scalar field contribution to metric field equations they remain unchanged and we will not comment on them in detail but instead refer the reader to \cite{Baumgarte:2012xy,Ruchlin:2017com}. Below we will discuss only the scalar field equation written within the considered formalism.

The $3+1$ decomposition of the scalar field equation, written with respect to the barred conformal metric and its covariant derivatives is given by
\begin{eqnarray}
(\partial_t - \beta^i \partial_i)\varphi &=& -\alpha K_\varphi , \\
(\partial_t - \beta^i \partial_i) K_\varphi &=& -e^{-4\phi} {\bar D}_i \alpha {\bar D}^i \alpha - \alpha \left[ e^{-6\phi} {\bar D}_i (e^{2\phi} {\bar D}^i) \right. \notag \\
&& \left. -K K_\varphi + \frac{1}{4} \lambda^2 \frac{df}{d\varphi} {\cal R}_{GB}^2 \right] ,
\end{eqnarray}
where the covariant derivatives $ {\bar D}$ are taken with respect of the conformal metric ${\bar \gamma_{ij}}$ and $K$ is the trace of the extrinsic curvature. For the solution of the above equations on top of the GR evolution we have developed an extension of the NRPy+ code \cite{Ruchlin:2017com}. This code allows to solve the GR field equations in the BSSN formalism in Cartesian or curvilinear coordinates. The latter choice speeds up the calculations considerably, since the head-on collision we consider possesses axial symmetry. 

In our calculations we have adopted a grid with resolution $400x64x2$ in the radial and the two angular directions. Taking only 2 points in the $\varphi$ direction reduces the evolution to axial symmetry, suited to our simulations. The two black holes collide starting from a separation of $5M$. That was close to the maximum we could achieve assuming a reasonable computation time and being able to observe the physically interesting phenomena. The reason why higher separation requires much larger resolution, especially in the $\theta$ direction, is the spherical grid implemented in {\tt NRPy+}. It is centered at the origin, which is not optimal in case of a large black hole separation. Even though an initial distance between the black holes of $5M$ is small if we want to be able to derive accurate waveforms of an actual merger, it is enough to observe the (de)scalarization of the black holes before and during the merger. Thus, it serves our goal well, which is to qualitatively study the effect of descalarization with a jump. In our simulations, we have also used Kreiss-Oliger numerical dissipation \cite{Kreiss1973} for the scalar field equations similar to \cite{Werneck:2021kch}, with a dissipation strength varying between 0.01 and 0.05 depending on the particular simulations, in order to avoid undesired high-frequency noise in the scalar field evolution. 

The extension of the NRPy+ code developed for the purpose of the present study was verified in the following way. We will comment only on the testing of the scalar field equation since the evolution of the pure Einstein equations remains unchanged and it has already been extensively discussed in \cite{Ruchlin:2017com}. First, we have confirmed that the scalar field has a 4th order convergence that is the same as for the evolution of the metric quantities. In addition, we have verified that the scalar charge of the newly formed black hole and the transition point from scalarized to GR solutions coincide within a few percents with the results from the scalar field evolution around isolated black holes \cite{Doneva:2021dqn}. 

\section{Black hole head-on collision}
We have considered the head-on collision of two equal mass black holes with mass $0.5M$ located at a distance of $5M$. As for the sGB theory, we have fixed $\beta=6$ in the coupling \eqref{coupling} and worked with two different values of $\kappa$, namely the case of standard scalarization $\kappa=0$ that is similar to the studies in \cite{Silva:2020omi}, and $\kappa=16$ for which the quartic term in the coupling \eqref{coupling} starts being important and a jump between the stable scalarized branch and Schwarzschild is observed as discussed in detail in the previous section. For both cases a number of simulations with different $\lambda/M$ values were performed that are in the region around $\lambda_{\rm min}$ and $\lambda_{\rm bif}$ in Fig. \ref{fig:D_lambda}. 

\begin{figure}
\includegraphics[scale=0.35]{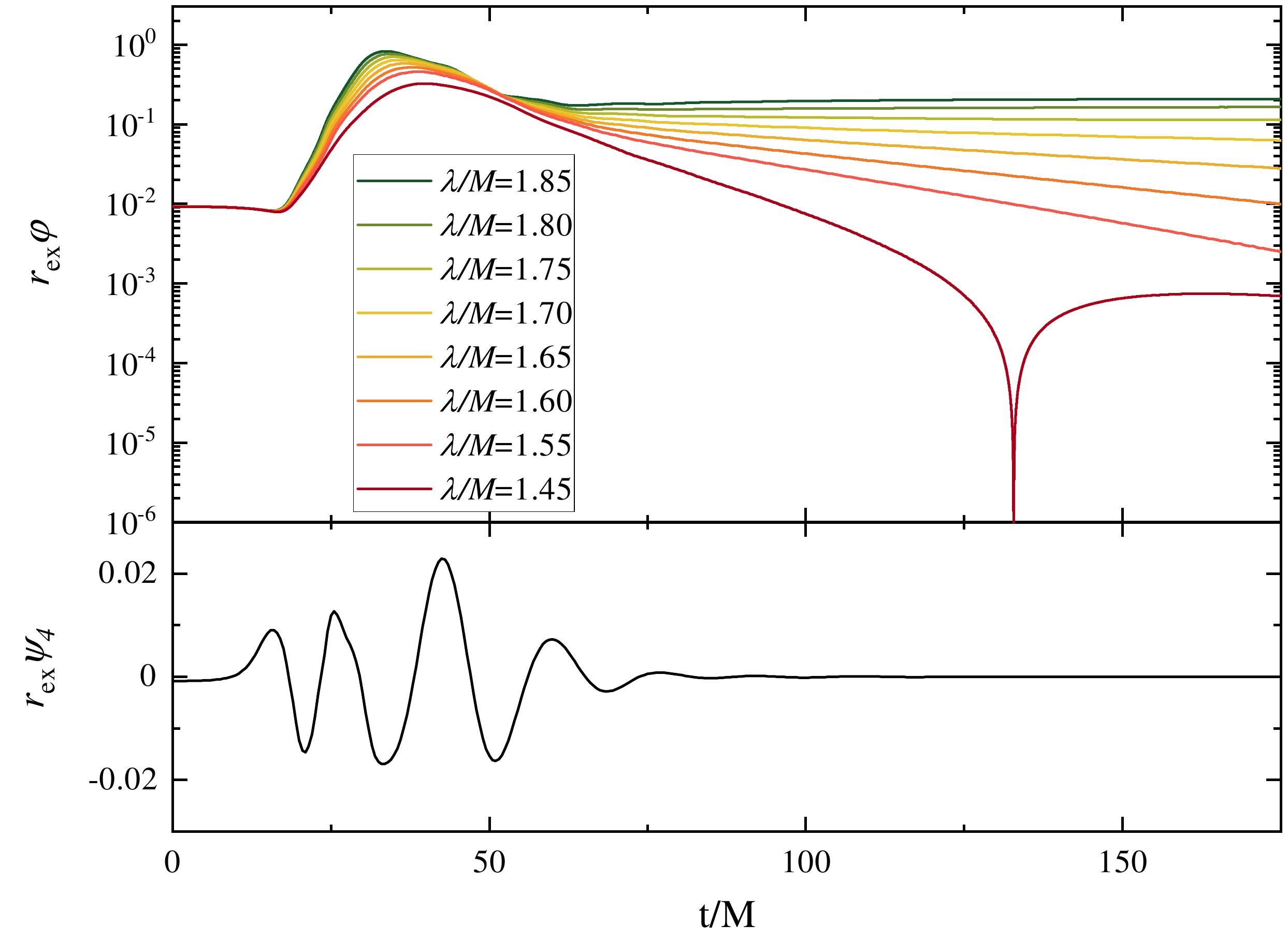}
\caption{(top) The evolution of the scalar charge $r_{\rm ex} \varphi$ as the head-on collision proceeds for $\beta=6$, $\kappa=0$ and several $\lambda/M$ close to the bifurcation point, where $r_{\rm ex}=12.5M$. (botton) Dominant ($l=2$, $m=0$) mode of $\psi_4$ times $r_{\rm ex}$ for a comparison, extracted at the same $r_{\rm ex}$.}
\label{fig:D_t_kappa0}
\end{figure}

\begin{figure}
\includegraphics[scale=0.35]{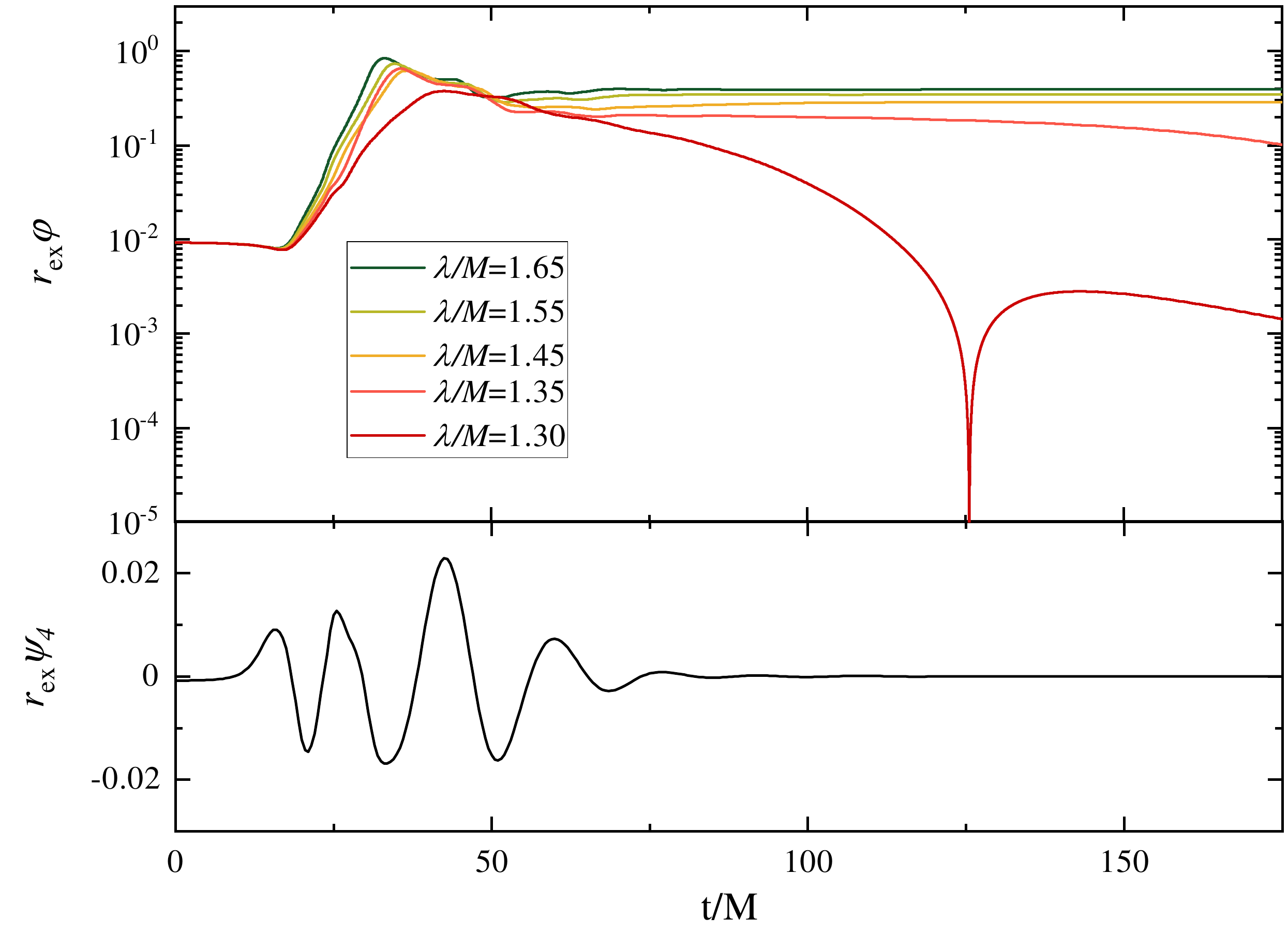}
\caption{(top) The evolution of the scalar charge $r_{\rm ex} \varphi$ as the head-on collision proceeds for $\beta=6$, $\kappa=16$ and several $\lambda/M$ close to the bifurcation point, where $r_{\rm ex}=12.5M$. (botton) Dominant ($l=2$, $m=0$) mode of $\psi_4$ times $r_{\rm ex}$ for a comparison, extracted at the same $r_{\rm ex}$.}
\label{fig:D_t_kappa16}
\end{figure}

The time evolution of the scalar charge for $\kappa=0$ is presented in Fig. \ref{fig:D_t_kappa0} while the $\kappa=16$ case is plotted in Fig. \ref{fig:D_t_kappa16}. In the lower panels of both figures the dominant $l=2$, $m=0$ mode of $\psi_4$ multiplied by the extraction radius is shown for comparison. Remember, that we work in the decoupling limit approximation that leaves the evolution of the metric quantities unchanged for any values of $\beta$, $\kappa$, and $\lambda$. The extraction radius is at $r_{\rm ex}=12.5M$ and we have verified that at this distance the scalar charge is already saturated to a constant with a relatively good accuracy. All values of $\lambda/M$ are chosen around $\lambda_{\rm bif}$ so that the two individual black holes with mass $0.5M$ are well within the scalarization window, while the newly formed black hole after the merger is on the border of scalarization.

In our simulations we used Brill-Lindquist initial data for the metric quantities. The scalar field evolution starts from a small perturbation and as time proceeds, the two individual black holes quickly develop scalar hair. In both Figs. \ref{fig:D_t_kappa0} and \ref{fig:D_t_kappa16} one can observe the exponential growth of the scalar field at early times. After a certain point, this exponential growth starts saturating. However, it cannot reach proper equilibrium due to the short time until the merger that is a consequence of the limited initial separation between the black holes. The evolution after the merger differs qualitatively for $\kappa=0$ and $\kappa=16$.

Let us first focus on the $\kappa=0$ case depicted in Fig. \ref{fig:D_t_kappa0}. As the black holes merge, the scalar field starts to decrease until it either reaches a new equilibrium for larger $\lambda/M$ or it starts decreasing exponentially for smaller $\lambda/M$ lower than $\lambda_{\rm bif}$. Close enough to the bifurcation point both the real and the imaginary parts of the scalar quasi-normal mode (QNM) frequency tend to zero and practically no oscillations can be observed in the case of descalarization and scalar field emission. Instead, one can see only a slow exponential decay for $\lambda/M$ smaller but close to $\lambda_{\rm bif}$. The well known form of the QNMs with oscillations and a subsequent tail is recovered only for $\lambda/M$ sufficiently smaller than $\lambda_{\rm bif}$ (e.g. $\lambda/M<1.35$ in Fig. \ref{fig:D_t_kappa0}).

It is clear that for $\kappa=0$ the transition between the two regimes of scalarized and non-scalarized black hole remnant is continuous with the scalar field being decreasingly small as the bifurcation point is approached. In addition, the small damping/growth time of the scalar field close to that point will lead to a continuous and gradient effect on the overall evolution of the system in case the full coupled system of the metric and scalar field evolution is considered. Thus, it might be difficult to discriminate between these two regimes in the gravitational wave data.

The picture changes qualitatively for large enough $\kappa$. From the results resented in Fig. \ref{fig:D_t_kappa16} one can easily see that there is an abrupt change of the scalar field evolution/emission as $\lambda/M$ passes below the minimum one $\lambda_{\rm min}$ for which scalarized solutions exist. Note that this point is different from the point of bifurcation $\lambda_{\rm min}<\lambda_{\rm bif}$. The main differences with the $\kappa=0$ case are the following. First, as $\lambda/M$ decreases and passes through $\lambda_{\rm min}$, the scalar charge changes with a jump that can have a clear signature in the GW observations of the merger itself and the subsequent QNM ringing. Even more importantly, for any value of $\lambda/M$ lower than $\lambda_{\rm min}$, the real and imaginary parts of the Schwarzschild scalar field QNM frequency are already substantially different from zero. This results in a rapid descalarization of the new BH in case $\lambda/M$ is lower than the scalarization threshold. If one considers the coupled evolution of the scalar field and the metric, then the rapid scalar field decrease that carries energy away to infinity will certainly have also influence on the emitted gravitational wave signal. Estimating this effect is beyond the scope of the present paper since we are considering the evolution in the decoupling limit approximation. If such information is available from full numerical simulations beyond the decoupling limit, though, and a large enough number of binary merging events are observed, one will be able to tell apart the different behavior of the system observed in Figs. \ref{fig:D_t_kappa0} and $\ref{fig:D_t_kappa16}$. This will naturally lead to strong constraints on the parameter space of the theory and even more generally -- it can potentially even completely discard certain types of couplings in sGB gravity.

Note that we have normalized all quantities with respect to $M$ for convenience. Since $\lambda$ is a dimensional parameter it can be also used for normalization and that has some advantages when discussing the observational implications below. In that case one can easily rescale the dependences in Figs. \ref{fig:D_t_kappa0} and \ref{fig:D_t_kappa16} and present the quantities in terms of $\lambda$ units. Then the different simulations for different $\lambda$ we have performed will be equivalent to simulations with a fixed $\lambda$ but different initial black hole masses. For example, the range of explored $\lambda/M$ in Figs. \ref{fig:D_t_kappa0} and \ref{fig:D_t_kappa16} can be translated to different initial masses of the merging black holes if the quantities are normalized with respect to $\lambda$. Taking intro account that we have worked with $M_b=0.5M$, the range of explored $\lambda/M$ in Figs. \ref{fig:D_t_kappa0}  translates to a mass range from $M_{\rm b\;min}/\lambda=0.270$ to $M_{\rm b\;max}/\lambda=0.345$, while in Fig. \ref{fig:D_t_kappa16}: from $M_{\rm b\;min}/\lambda=0.303$ to $M_{\rm b\;max}/\lambda=0.385$. In practice, a fixed sGB theory means fixing the parameter $\lambda$ to a specific value and multiple gravitational wave observations will provide us with the opportunity to observe mergers with different black hole masses. In the discussion below we will consider exactly this scenario.

\section{Discussion and observational prospects}
We have performed simulations of black hole head-on collision of equal mass black holes in sGB gravity observing the process of descalarization. The scalar field dynamics is explored on the top of the GR background, thus neglecting the scalar field backreaction. This allows us to capture qualitatively the main features of descalarization and make predictions for the possible observational manifestations. In contrast to previous works, we have focused on the case when the coupling function between the scalar field and the Gauss-Bonnet invariant has both quadratic and quartic terms in the scalar field. This changes the picture of scalarization not only quantitatively but also qualitatively. Namely, the branch of stable scalarized solutions is not continuously connected to Schwarzschild and there is a jump between both. In addition, there is a region of the parameter space where both the GR and the scalarized branches are stable. 

This property of the solutions leads to a very interesting phenomenology of the black hole mergers. In order to demonstrate it, we have considered a set of parameters for which the individual binary black holes are well within the scalarization window. Thus, any small scalar field perturbation will lead to rapid development of a nontrivial scalar hair. The newly formed black hole, though, is on the border of scalarization. Namely, for a fixed coupling parameter $\lambda$ and as the black hole mass is varied, we make a transition between the regimes where the resulting black hole with mass $M_f/\lambda$ is either scalarized or $M_f/\lambda$ is large enough and only the GR solution exists. The limiting value of the mass dividing the two regimes is different for the case of standard scalarization with zero quartic term and $\kappa=0$ (denoted by $M_{\rm bif}$ that coincides with the point of bifurcation of the scalarized branch from Schwarzschild) and the case when the quartic term is strong enough with $\kappa>0$ (denoted by $M_{\rm max}>M_{\rm bif}$).

We clearly observed that for $\kappa=0$ the scalar charge of the resulting black hole after the merger goes to zero continuously as $M_f/\lambda$ approaches $M_{\rm bif}$. If we thus observe a number of merger events with a dense distribution of their total mass, and take a series of such events with increasing total mass, we will observe the following. The newly formed black hole will slowly decrease its scalar charge as the mass $M_f$ increases. At a certain point the initial binary masses will become such that $M_f$ surpasses $M_{\rm bif}$ that will result in a descalarization after the merger. For $M_f$ close to the bifurcation point $M_{\rm bif}$ the scalar charge is very weak and it will have a small influence on the black hole dynamics after the merger. Moreover, the damping time of the scalar field tends to zero at the bifurcation point leading to a slow emission of the remaining scalar field. Such a slow energy release is unlikely to have a strong influence on the binary dynamics and GW emission. That is why the transition between the two regimes before and after $M_{\rm bif}$ will not be easily seen in the GW signal. Of course, the presence of scalar charge of the individual merging black holes can be detected through other methods, especially in the inspiral phase of an actual merger (see e.g. \cite{Khalil:2019wyy}).

For a strong enough quartic term in the coupling with $\kappa>0$ there is a jump between the last stable scalarized solution and Schwarzschild. That is why we observed that the scalar charge of the merger remnant saturates to a fixed value with the increase of the binary mass $M_f$ until the threshold $M_{\rm max}$ is reached and it jumps to zero. More specifically, the observed merger remnants for a series of observations will be divided into two parts -- merger remnants with a strong scalar field that clearly has a different ringdown compared to GR, and a GR merger remnant. There will be no continuous transition between both since there are no stable intermediate weak scalar field black holes. In addition, the Schwarzschild black hole with $M_{\rm max}$ is already stable and has a relatively short damping time of the scalar gravitational radiation resulting in a rapid emission of the scalar field that has developed before the merger. Thus, if one observes a large number of binary mergers with a sufficient density in the mass distribution, such a jump can be clearly observed in the GW signal.

The process described above has very interesting similarities with the first-order matter phase transition from confined hadronic to deconfined quark matter. In that case, as well it happens that if the initial mass of the merging neutron stars surpasses a given threshold, a phase transition can happen during the merger resulting in a newborn supramassive neutron star with a quark matter core. This process was first simulated in \cite{Most:2018eaw,Bauswein:2018bma,Weih:2019xvw} demonstrating the abrupt change in the merger characteristics in the presence of such phase transition. The methodology developed for searching of such matter phase transitions in the GW signal can be readily applied for black hole mergers in sGB gravity since the process shares interesting similarities.

We should keep in mind that in the original sGB theory the scalar gravitational radiation is not directly coupled to the perturbations of the metric and is thus directly not observable with the gravitational wave detectors. The scalar waves carry away energy, though, that will leave clear imprints on the observed gravitational wave signal. For example, there will be a phase difference between a merger of GR black holes and the scalarized ones, with the latter ones merging faster because of the accelerated inspiral. In the case of dynamical descalarization, it might happen that such scalar wave energy loss is present during the inspiral but the newly formed black hole is just Schwarzschild that will clearly lead to specifics in the data analysis of both the inspiral and the ringdown phase. A rapid release of scalar energy during merger in case of descalarization can also alter the dynamics of the actual merger if the backreaction of the scalar field on the spacetime dynamics is taken into account. Last but not least, the sGB gravity can be slightly modified to include such a direct coupling between the scalar and metric perturbations leading to the presence of potentially observable breathing modes. 

Even though we stick to the sGB gravity in the present paper, the conclusions we have made are more or less independent on the particular coupling and the theory itself, as long as we have a black hole scalarization mechanism and a coupling allowing for a gap between the stable solutions. The observation of an abrupt change in the gravitational wave signature happening at a certain black hole mass will be a hint of the presence of the effect described above. On the other hand, if enough gravitational wave observations are accumulated without clear evidence for the presence of such a jump, a whole class of couplings can be excluded.  

We should note that the binary pulsar experiments have already constrained the black hole scalarization in sGB gravity quite strongly \cite{Danchev:2021tew} allowing for scalarized black holes only below ten solar masses. We should note that the studies in \cite{Danchev:2021tew} were performed only for one coupling with $\kappa=0$ and for massless scalar fields. It can well happen that for different couplings or a modification of the original sGB theory such as \cite{Antoniou:2022agj} the picture changes. Moreover, even if a tiny scalar field mass is present it can suppress the scalar dipole radiation from the pulsars in close binary systems thus reconciling the theory with observations. 

The future prospect is to first consider a more realistic merging process, i.e. binary inspiral instead of a head-on collision. In addition, the decoupling approximation should be dropped in order to be able to obtain a full picture of the manifestations of the described effect in the emitted gravitational wave signal.

\section*{Acknowledgements}
DD  acknowledges financial support via an Emmy Noether Research Group funded by the German Research Foundation (DFG) under grant no. DO 1771/1-1.  SY would like to thank the University of Tuebingen for the financial support. The partial support by the Bulgarian NSF Grant KP-06-H28/7 is acknowledged. AVV thanks FCT for financial support through Project~No.~UIDB/00099/2020.


\bibliographystyle{unsrt}
\bibliography{references}

\end{document}